# A SCALE-INVARIANT LOCAL DESCRIPTOR FOR EVENT RECOGNITION IN 1D SENSOR SIGNALS

*Jierui Xie*[1], *Mandis S. Beigi*[2]

[1] Department of Computer Science, Rensselaer Polytechnic Institute, Troy, NY 12180
[2] IBM TJ Watson Research Center, 19 Skyline Drive, Hawthorne, NY 10532
Email: xiej2 @cs.rpi.edu, mandis@us.ibm.com

**ABSTRACT**

In this paper, we introduce a shape-based, time-scale invariant feature descriptor for 1-D sensor signals. The time-scale invariance of the feature allows us to use feature from one training event to describe events of the same semantic class which may take place over varying time scales such as walking slow and walking fast. Therefore it requires less training set. The descriptor takes advantage of the invariant location detection in the scale space theory and employs a high level shape encoding scheme to capture invariant local features of events. Based on this descriptor, a scale-invariant classifier with "R" metric (SIC-R) is designed to recognize multi-scale events of human activities. The R metric combines the number of matches of keypoint in scale space with the Dynamic Time Warping score. SIC-R is tested on various types of 1-D sensors data from passive infrared, accelerometer and seismic sensors with more than 90% classification accuracy.

## 1. INTRODUCTION

In this paper, "*multi-scale*" events denote a collection of events belonging to the same semantic class which may take place over different time durations. For example, "*walking*" describes a multi-scale collection of events which may include "*walking slowly*", "*walking briskly*", "*walking fast*", etc.

With only few training data, most classification models suffer from the problem of over-fitting and are sensitive to fluctuations in data. They perform even worse when the event of interest is multi-scale. This paper introduces a solution to this challenge. Our work has been inspired by Scale-Invariant Feature Transform (SIFT) [1, 2], which is originated in scale space theory where it was designed to extract local features from images. The features are invariant to image scale and rotation. However, SIFT is tailored to deal with images, while we are interested in detecting events from 1-D time series sensor signals (e.g., passive infrared, accelerometer or seismic sensors) in which image intensity related concepts such as orientation are not applicable. The closest work with ours is [3], which applies SIFT to 1D circular image and uses both color information and curvature as the descriptor.

In this paper, we introduce a scale invariant feature descriptor for event recognition from sensors. We assume that while a multi-scale event can evolve and last over various time periods, each event will still generate some relatively invariant "local features". Our objective is to find and describe these features of the events of interest, and then to use such features to recognize other instances of the same event class over multiple time scales in unknown signals.

## 2. SCALE-INVARIANT FEATURES DESCRIPTOR

In our experiments, the signal is 1-D discrete time series sampled at a certain sampling rate depending on the sensor type. Each individual time series is segmented manually from a sensor stream such that it contains only one event.

A scale-space keypoint detector is first employed to identify the potential stable locations (as known as "keypoints") that may be robustly identified over multiple time scales. Then, local feature vectors are extracted using our proposed shape based descriptors.

### 2.1. Keypoint detection

We adopt the Differences of Gaussians (DoGs), a scale-invariant keypoint detector proposed by Lowe [2], and tailor it for 1-D sensor signal. The scale space is defined as a function $L(x, \sigma)$, produced from the convolution of a variable-scale Gaussian $G(x, \sigma)$, with a discrete time input signal $I(x)$.

$$L(x,\sigma) = G(x,\sigma) * I(x) = \frac{1}{\sigma\sqrt{2\pi}} e^{-\frac{x^2}{2\sigma^2}} * I(x)$$

Where "*" is the convolution operation. First, the initial signal $I(x)$ is convolved with a Gaussian function $G_0$ with $\sigma_0$. resulting in a "blurred" signal $L_0$. $L_0$ is used as the first signal in the Gaussian pyramid and is incrementally convolved with a Gaussian $G_i$ using $\sigma_i$ to create the $i^{th}$ signal $L_i$ which is equivalent to the original signal filtered with a Gaussian $G_k$ with $k\sigma_0$. Following the practice in [1], $\sigma_i$ is set to $1.6k$ and increased by a user specified constant multiplicative factor $k=2^{1/4}$. The upper bound of $\sigma$ and thus the number of scales is determined by the length of input signal. The DoG function, $D(x, \sigma)$, is computed from $L$'s two adjacent scales: $D_k(x, \sigma) = L(x, k\sigma)-L(x,(k-1)\sigma)$ (Fig. 1(a)).

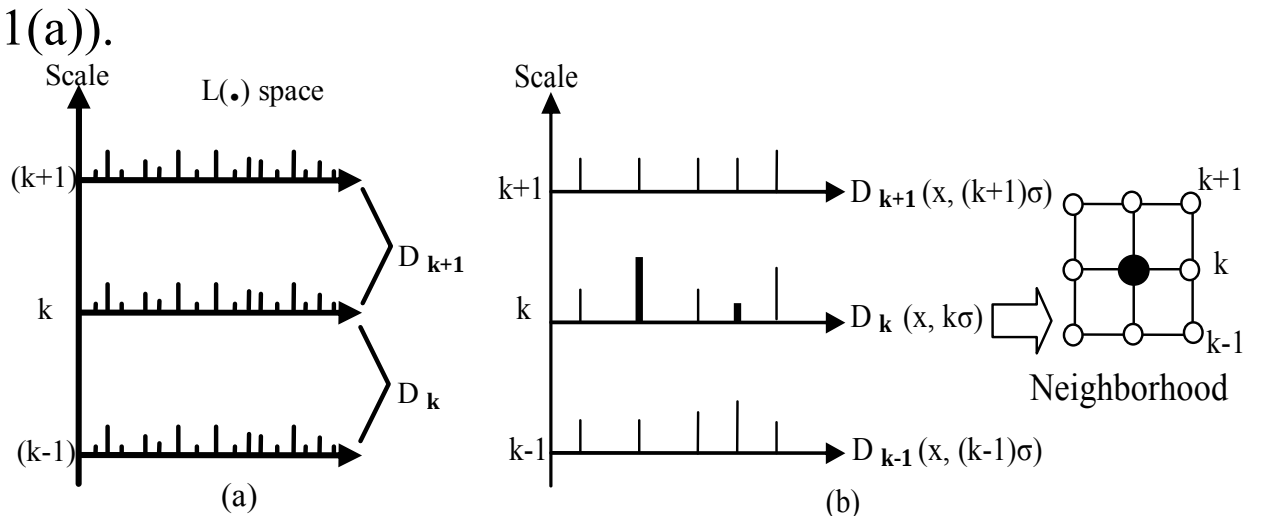

Fig.1: The calculation of DoG. $D_k=L(x,k\sigma)-L(x,(k-1)\sigma)$ $D_{k+1}=L(x,(k+1)\sigma)- L(x,k\sigma)$; (b) The illustration of extrema at scale $k$(bold points) and the neighborhood (white points) of the extrema (black point)

The keypoints are defined as the local extrema (i.e. minima and maxima) of $D_k(x,\sigma)$ by comparing each sample point in every scale to its two neighbors (Fig. 1(b)) in the same scale $k$ and three neighbors in the scale below $(k-1)$ and three neighbors in the scale above $(k+1)$ having a total of 8 neighbors. If the point is minimum or maximum, it is considered to be an extrema point.

### 2.2. Feature descriptor

After the keypoints of an event signal are located, feature descriptors are created to represent the "local property" of the signal for each keypoint. The descriptor attempts to capture how "hilly" the signal around the keypoint is and what its characteristics are by encoding the important shape information. Instead of from raw data, high level features are extracted from the shapes such as peaks and slopes, which are expected to be more robust to scale change and noise. Each descriptor is a vector that describes a small region around the keypoint. The sequence of descriptors forms a scale invariant pattern of the signal.

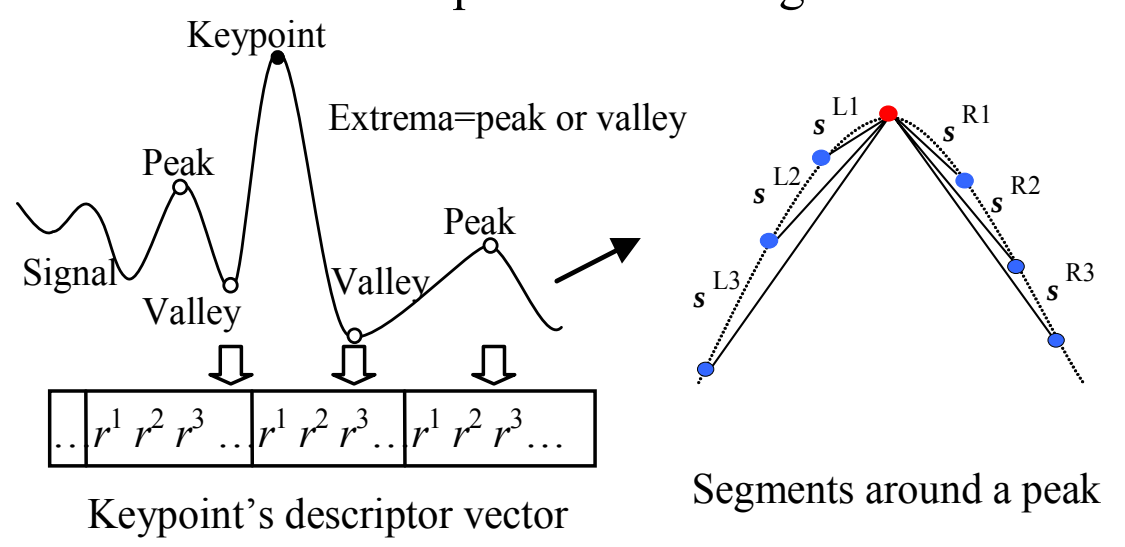

Fig. 2: Illustration of keypoint descriptor vector construction

An *extrema* is defined as either a peak or a valley which passes the *monotonicity test*, which tests whether the values of consecutive neighboring points are monotonically increasing or decreasing, and then only accepts extremas with large enough neighborhood (e.g. distance of 20 points in our experiments). The test removes the impact of smaller sudden changes or noise.

Given a keypoint and parameters $N$ and $M$, where $N$ is the number of extrema around a keypoint and $M$ is the number of segments on either side of each extrema (see Fig. 2). The descriptor is created as follows:

1. Find $N$ nearest extrema points around the keypoint, i.e. $N/2$ extremas to the left and $N/2$ extremas to the right of the keypoint.
2. Encode the characteristics of each extrema point as a vector in the following steps: (a) take $M$ nearest neighboring points to the left and $M$ nearest neighbors to the right of the extrema point. The first points on the left and the right side are labeled as $p^{L1}$ and $p^{R1}$, respectively, $p^{L2}$ and $p^{R2}$ *for the second* pair, and so on. We then calculate the slopes of segments *[extrema* $p^{Li}$*]* and *[*$p^{Ri}$ *extrema]*, i.e., $S^{Li}$ and $S^{Ri}$ (see Fig. 2); (b) calculate the ratio of these two slopes, $r^i = S^{Ri}/ S^{Li}$; (c) if the extrema is a peak, take the absolute value of $r^i$; (d) save $r^i$ as an element of the descriptor vector, and repeat step 2 for all the $M$ pairs of extrema points.

The resulting descriptor, $[r^1, r^2, r^3,...,r^{NM}]$, is a vector of size $N*M$ per keypoint. $N$ and $M$ are configurable parameters that essentially determine how large the local region should be considered. (In our experiments, $N=M=4$ for the seismic signal that samples at the rate of 8159 samples/second).

The ratio of the slopes, $r^i$, for either a peak or a valley has a negative value because the slopes on either side have opposite signs. In Step 2(b), we make the $r^i$ for peaks to be positive. This results in a penalty for opposite direction extremas when calculating matching distance. This penalty improves the distinctiveness of the descriptor.

The shape of an extrema point is expressed by *caps* (formed by two solid lines as shown in Fig. 3(a)), which are formalized by the slope ratio $r^i$. We assume that the patterns of multi-scale events tend to keep similar shapes over different time scales, i.e. changes in shape are proportional in most of the time scales. Both solid lines of the cap either go outward or inward (either enlarging or shrinking). Our descriptor essentially punishes disproportionate changes (i.e. changes in any single direction in Fig. 3(b)) while measuring the similarity between two signals.

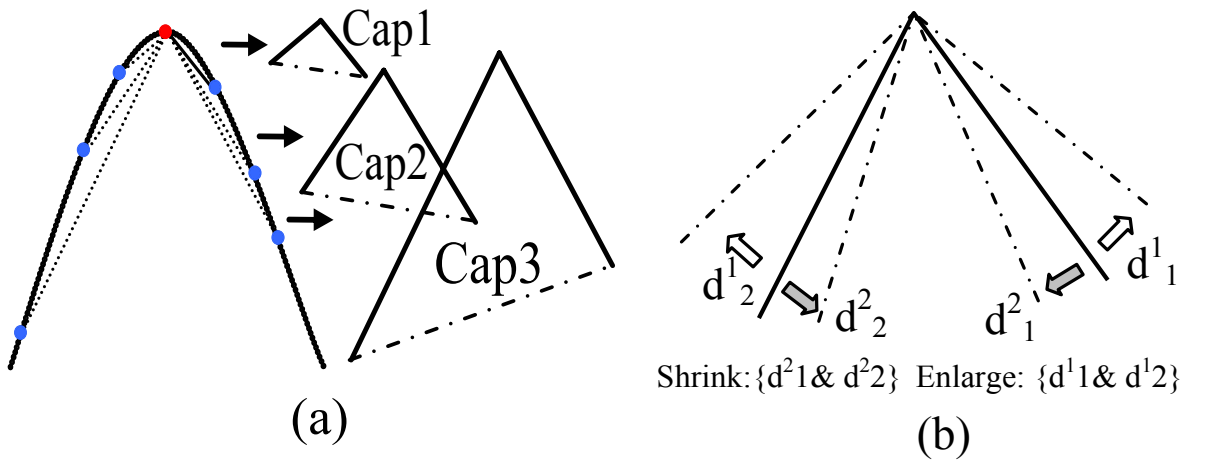

Fig. 3: The caps and change directions

For two events occurring at different time scales (scale 1 and 2 in Fig. 4), if scale2 is proportional to scale1 (i.e. it either enlarges or shrinks features of scale 1 by a constant factor $s$), then the ratio $r=r_2^i/r_1^i$ (where $r_1^i = -\cot(\theta_2)/\cot(\theta_1)$ corresponding to scale1 and $r_2^i = -\cot(s\theta_2)/\cot(s\theta_1)$ corresponding to scale2) is expected to be *close*. We set consider $r \in [2/3, 3/2]$ as *close* in which the upper bound is set to 3/2 the same as SIFT's matching threshold. Since signals in different time scales are unlikely to rotate (peaks cannot become opposite when the scale changes), for a peak $\theta_1, \theta_2 \in [0, \pi/2]$, $s*\theta_1$ and $s*\theta_2$ are also in [0, $\pi/2$]. Based on this observation, we conduct a test using the following parameters at the given ranges:

$\theta_1 = [0.05 : 0.005 : 0.45]\pi$

$\theta_2 = [\theta_1 : (0.45*\pi - \theta_1)/100 : 0.45*\pi]$

$s = [1.25 : 0.05 : (0.45*\pi/\theta_2)];$

The result shows that there are 33462 out of 47329 (70.7%) cases in which $r$ is *close*. This indicates that in at least 70.7% of the time, the descriptor will have high accuracy. The same analysis also applies to *valley*.

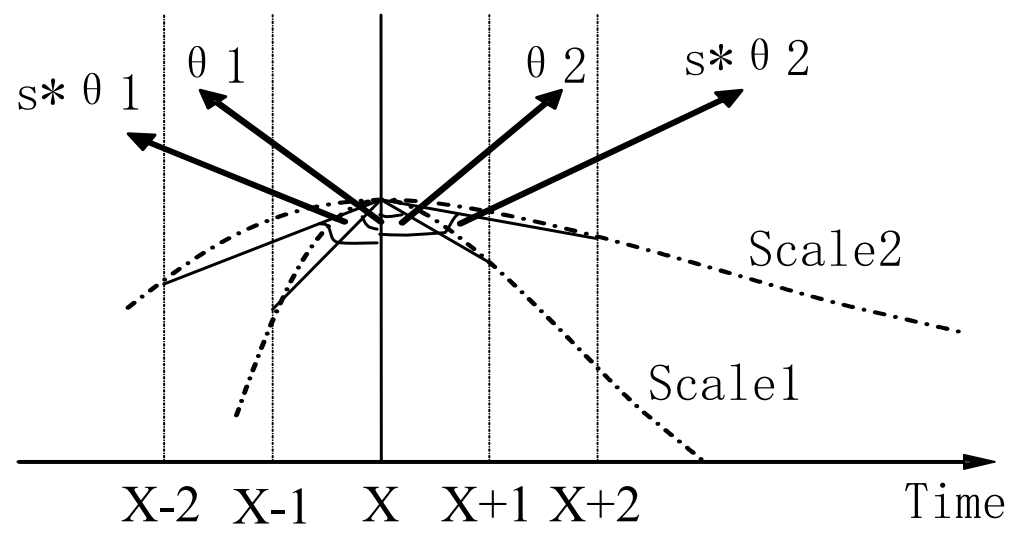

Fig. 4: Shape-based feature descriptor

### 2.3. Keypoint matching in multi-time scale events

To show how the above ideas apply to the event signals captured by sensors, we evaluate the match between various pairs of multi-time scale signals. The matching between two signals is measured by the keypoint matching of two signals. Nearest neighbors (NN) matching and DTW (Dynamic Time Warping [5] are two widely used sequence matching techniques. We apply NN-Matching in this section and DTW later on.

**NN-Matching**: Firstly, the descriptors (keypoints) of pattern signal and query signal are calculated. For each descriptor of the pattern signal, we calculate all the pair-wise Euclidean distances from it to all the descriptors from query signal. Secondly, we compare the nearest and second nearest match. Only if the difference between these two is larger than a certain threshold (we set nearest/second nearest=1.5), the matching with smallest distance is claimed to be the candidate match between this keypoint with the query signal. After matches of all descriptors of the pattern signals are calculated, we apply RANdom SAmple Consensus (RANSAC) [4] to remove the false candidate matches from a global view. After using RANSAC, the remaining matches and corresponding keypoints represent the invariant pattern in these two signals. The number of matches is called the "*M* metric".

The NN-Matching is applied to multi-scale activities from Passive Infrared (PIR) sensors. It includes *walking* at normal pace, very slowly, briskly, and fast. "*Walking normal*" serves as the pattern signal. The other three types of walking are query signals. Fig. 5 shows the matching results of the two of those experiments. As shown, the shape properties (e.g., slope) around the corresponding matched descriptors over different temporal scale events are highly similar. It indicates that the descriptors are able to catch some significant local patterns of the events, which are somehow repeated in multi-scale instances.

Note that Fig. 5 also shows keypoints that are not peaks or valleys. This happens because the keypoints locations in the figure are the mapped view from Gaussians space (containing multiple scales) to the raw signal coordination (temporal). The keypoints in its own scale are always a peaks or valleys.

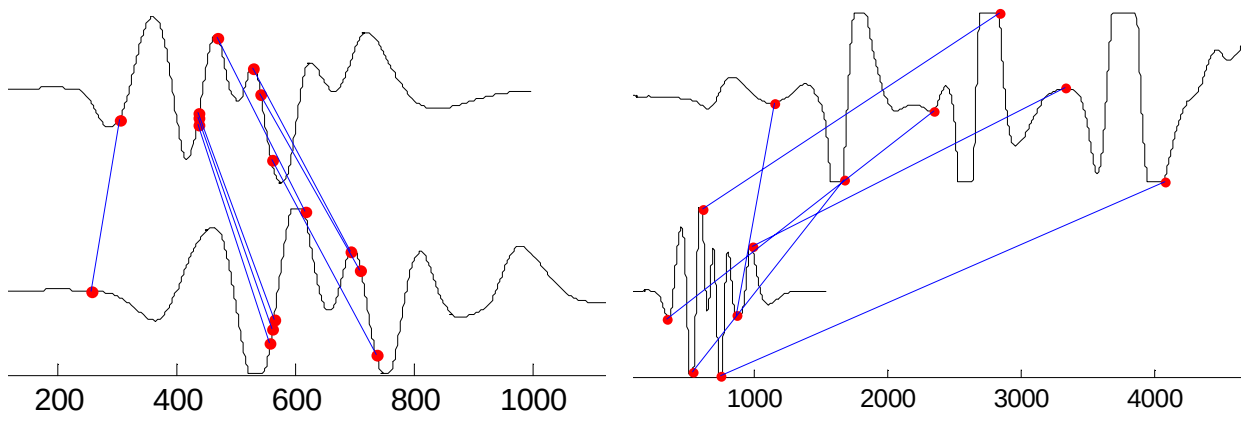

Fig. 5: Matching between multi-scale walking from PIR sensor. The lines connecting pairs of descriptors from two signals indicate the matching. Left: walking normal (bottom) vs. walking briskly (top); Right: walking normal (bottom) vs. walking slowly (top)

## 3. CLASSIFIER AND DISTINCTIVENESS EVALUATION

### 3.1 The framework of the classifier

Since the descriptors are able to encode the local invariant features, they can serve as good features. Based on these descriptors, we build Scale Invariant Classifier (SIC) to perform event recognition. The general framework of SIC includes "keypoints detection", "shape-based feature descriptors construction", "keypoint matching", "metric calculation" and "classification output". Besides $M$ and $DTW$ score, we introduce a new score metric defined as $R = M / DTW$. Depending on the distance metric, the classifier is referred to SIC-M, SIC-D or SIC-R.

During the classification, one or more signals from each class are required. First, the query signal matches with every signal in the training set. Then, the average matching scores over each class are computed. These average scores represent the matching between the query signal and classes. Finally, the query signal is classified to the class with the highest score.

### 3.2. Invariance analysis on the metric R and SIC-R

We test the performance of SIC-R, SIC-M and SIC-D on 54 human activities from accelerometer sensors attached to the human body. This data set consists of 9 classes (walk, jump, kick, go-upstairs, go-downstairs, spin, squat, turn-left, turn-right) and each class contains 6 instances (total of 54 activities). As shown in Fig. 6, SIC-R, with about 95% classification accuracy (51 out of 54), performed much better than SIC-M (33%, 18 out of 54) or SIC-D (11%, 6 out of 54).

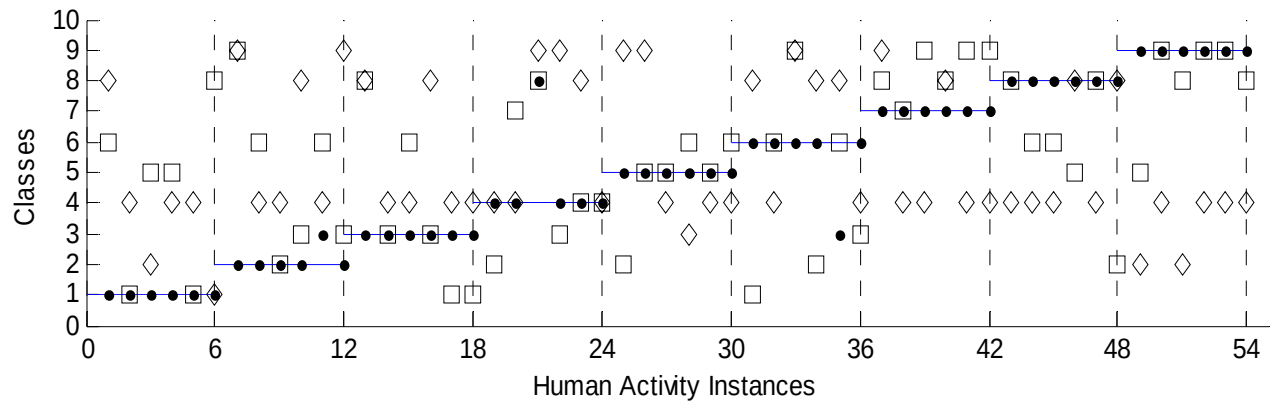

Fig. 6: The classification results of SIC-R (dot), SIC-M (square) and SIC-D (diamond) metric. The straight lines denote the expected classification. The symbols falling in the lines indicate being correctly classified.

The satisfying performance of "*R*" is due to the special characteristics of the keypoints over Gaussian space and multi-scale events. First, the detected keypoints are not perfectly ordered, i.e., the order of the keypoints from a certain scale can not be preserved in the raw signal coordination. Second, a certain location in the raw signal coordination may have more than one mapped points. Moreover, the multi-scale events have the following features: (a) of various lengths; (b) although longer signals tend to have more matching, the number of matching is unpredictable. These features make single commonly used measure like *M* metric (number of matches), or *DTW* not suitable. In practice, the normalization of *M* and *DTW* reduce the impact of variable length of the signal. However, *M* still does not perform well due to the use of Euclidean distance, which is not always appropriate [5] and *DTW* score is sensitive to the order of keypoints. Instead, R reduces the impact of variable signal length and non-perfect order still takes matching quality into account.

### 3.3. Comparison with HMM

We compare SIC-R with supervised classifier HMM. One HMM is built for one class of events. All HMMs adopt full-connected structures and take the optimal number of states (4 in our experiments). Baum-Welch algorithm is used for training. The performances are tested on the same dataset in 3.2..

SIC-R outperforms HMM with same size of training set, in witch *94.4% (51/54)* accuracy for SIC-R and *14.8%(8/54)* accuracy for HMM. Fig. 7 and Fig. 8 show the similarity matrix. The X axis is the activity id and Y axis is the group id. From the column's point of view, the color indicates the possibility of a certain activity being in different groups. The whiter the more likely. As shown in Fig. 7, SIC-R misclassified only three activities, i.e. 11, 21 and 35. Moverover, only one white color in most columns indicate strong distinctiveness between groups. Instead, HMMs only correctly classify 8 activities, i.e.10, 12, 24, 30, 34, 46, 52, and 54. With low accuracy, most columns of similarity matrix of HMM contain close colors, indicating the HMMs are not distinctive.

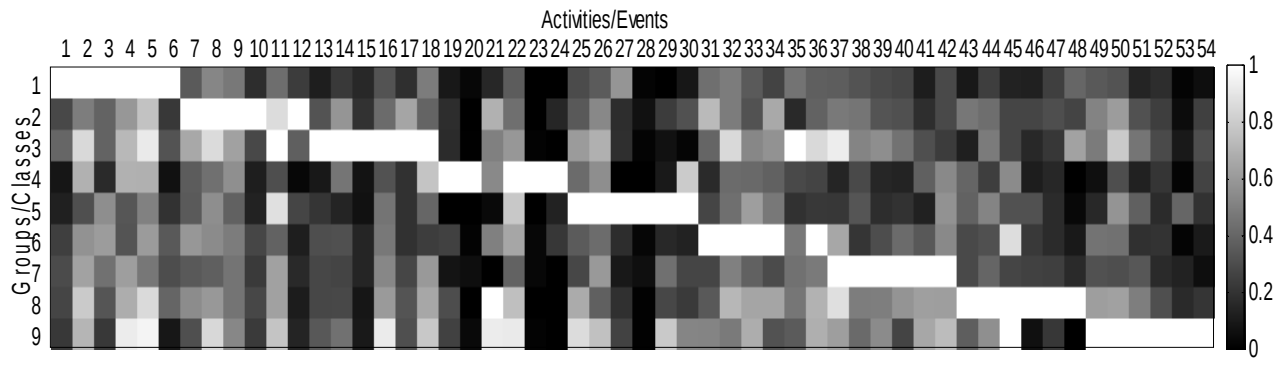

Fig. 7: Similarity matrix of SIC-R for Dataset 2

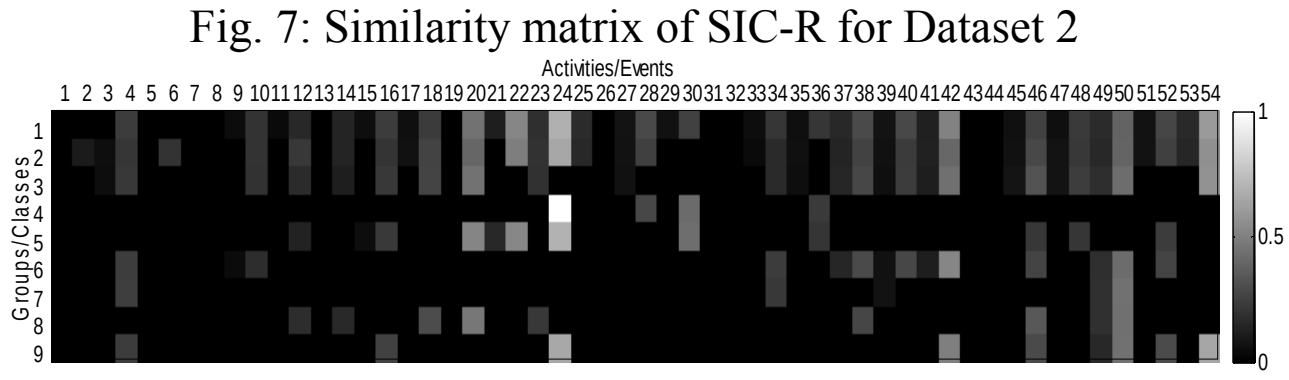

Fig. 8: Similarity matrix of HMMs for Dataset 2

## 4. CONCLUSION

We have proposed a scale invariant feature descriptor for 1D sensor signals based on the local "shape" information of the event's signal. The experiments use various kinds of sensors show that our classifier is discriminating and invariant. Our future work will focus on further improvement of the distinctiveness of the descriptors.

## 5. ACKNOWLEDGEMENT

Research was sponsored by the U.S. Army Research Laboratory and the U.K. Ministry of Defense and was accomplished under Agreement Number W911NF-06-3-0001. The views and conclusions contained in this document are those of the author(s) and should not be interpreted as representing the official policies, either expressed or implied, of the U.S. Army Research Laboratory, the U.S. Government, the U.K. Ministry of Defense or the U.K. Government. The U.S. and U.K. Governments are authorized to reproduce and distribute reprints for Government purposes notwithstanding any copyright notation hereon.

## 6. REFERENCES

[1] D.G. Lowe, "Distinctive Image Features from Scale-Invariant Keypoints", International Journal of Computer Vision, 60, 2, pp. 91-110, 2004.

[2] D.G. Lowe, "Object recognition from local scale-invariant features," Proceedings of ICCV, pp. 1150-1157, 1999.

[3] A. Briggs, Y. Li, and D. Scharstein, "Feature matching across 1d panoramas". 6th Workshop on Omnidirectional Vision, 2005.

[4] M.A. Fischler and R. C. Bolles, "Random Sample Consensus: A Paradigm for Model Fitting with Applications to Image Analysis and Automated Cartography". Comm. of the ACM 24: 381-395, 1981.

[5] D.Berndt and J. Clifford, "Using dynamic time warping to find patterns in time-series", AAAI-94 Workshop on Knowledge Discovery in Database, 1994.